\documentclass[a4paper,11pt]{article}

\usepackage{jheppub} 

\usepackage[T1]{fontenc} 
\usepackage{graphicx}
\usepackage{amsmath}
\usepackage{amssymb}

\def\U1mt{U(1)_{L_\mu-L_\tau}}
\def\wt{\widetilde}
\def\ol{\overline}
\def\nl{\nonumber\\}

\title{\boldmath Dark matter contribution to $b\to s \mu^+ \mu^-$ anomaly in local $U(1)_{L_\mu-L_\tau}$ model}

 \author{Seungwon Baek}
 \affiliation{School of Physics, KIAS, Seoul 02455, Korea}

\emailAdd{swbaek@kias.re.kr}

\abstract{We propose a local $U(1)_{L_\mu-L_\tau}$ model to explain $b \to s \mu^+ \mu^-$ anomaly observed at the LHCb
  and Belle experiments. The model also has a natural dark matter candidate $N$. We introduce $SU(2)_L$-doublet
colored scalar $\widetilde{q}$ to mediate $b \to s$ transition at one-loop level. The $U(1)_{L_\mu-L_\tau}$ gauge
symmetry is broken spontaneously by the scalar $S$. All the new particles are charged under $U(1)_{L_\mu-L_\tau}$. 
We can obtain $C_9^{\mu,{\rm   NP}} \sim -1$ to solve the $b \to s\mu^+\mu^-$ anomaly and can explain the correct dark
matter relic density of the universe, $\Omega_{\rm DM} h^2 \approx 0.12$, simultaneously,
while evading constraints from electroweak precision tests, neutrino trident experiments and other quark flavor-changing
loop processes such as $b \to s \gamma$ and $B_s-\overline{B}_s$ mixing. Our model can be tested by searching for $Z'$
and new colored scalar at the LHC and $B \to K^* \nu \overline{\nu}$ process at Belle-II.}

\begin{document} 
\maketitle
\flushbottom

\section{Introduction}
\label{sec:intro}
Flavor changing neutral current (FCNC) processes are sensitive probe of new physics (NP) beyond the
standard model (SM), of which the $b \to s \mu^+ \mu^-$ has drawn much interest recently due to
anomalies observed at the LHCb and Belle experiments.
A form-factor independent angular observable $P'_5$~\cite{DescotesGenon:2012zf} in the decay $B^0 \to K^{*0} \mu^+ \mu^-$ shows
3.7$\sigma$ discrepancy in the interval $4.3 < q^2 < 8.68$ GeV$^2$, $q^2$ being dimuon invariant mass squared~\cite{Aaij:2013qta}.
A global analysis of the CP-averaged angular observables indicates differences from the SM predictions at the level 
of 3.4$\sigma$~\cite{Aaij:2015oid}. The $P'_5$ anomaly has also been found by Belle collaborations at the
level of $2.1-2.6 \sigma$~\cite{Abdesselam:2016llu, Wehle:2016yoi}. 
For the $B_s^0 \to \phi \mu^+ \mu^-$ mode, the differential branching fraction has been found to be more than 3$\sigma$
below the SM predictions in the range $1 < q^2 <6$ GeV$^2$~\cite{Aaij:2015esa}. 
Similar tendency has been observed in $B \to K^{(*)} \mu^+ \mu^-$~\cite{Aaij:2014pli, Aaij:2016flj,Khachatryan:2015isa}, 
$\Lambda_b^0 \to \Lambda \mu^+ \mu^-$~\cite{Aaij:2015xza}.

Most interesting observables are the ratios~\cite{Hiller:2003js}
\begin{align}
R_{K^{(*)}} & \equiv \frac{{\cal B}(B \to K^{(*)} \mu^+ \mu^-)}{{\cal B}(B \to K^{(*)} e^+ e^-)},
\end{align}
which are predicted to be $1+O(m_\mu^2/m_b^2)$, representing the lepton-flavor universality (LFU) in the SM.
They are theoretically very clean because the hadronic uncertainties are canceled in the ratios. The measured value
$R_K$ at LHCb in the range $1 <q^2 < 6$ GeV$^2$ is $0.745^{+0.090}_{-0.074}({\rm stat}) \pm 0.036 ({\rm syst})$, deviating from
the SM predictions by $2.6\sigma$~\cite{Aaij:2014ora}. Recently the LHCb also measured $R_{K^*}$ with the results~\cite{Aaij:2017vbb}.
\begin{align}
R_{K^*} &= \left \{
\begin{array}{ll}
0.66^{+0.11}_{-0.07}(\rm stat) \pm 0.03 (\rm syst)  & {\rm for} \, 0.045 < q^2 < 1.1 \, {\rm GeV}^2, \nonumber\\
0.69^{+0.11}_{-0.07}(\rm stat) \pm 0.05 (\rm syst)  & {\rm for} \, 1.1 < q^2 < 6.0 \, {\rm GeV}^2, 
\end{array}
\right.
\end{align}
showing deviations at the level of $2.1-2.3 \sigma$ and $2.4-2.5 \sigma$  in the two $q^2$ regions, respectively.

It is worth noting that these possible deviations are in the same direction, and when combined, the discrepancy with the
SM predictions is at the level of $\sim5\sigma$~\cite{Descotes-Genon:2013wba,Altmannshofer:2017fio, Capdevila:2017bsm,
  Alok:2017jaf, Ciuchini:2017mik,Alok:2017sui}.
The $b \to s \ell^+ \ell^-$ decay is described by the effective weak Hamiltonian
\begin{align}
{\cal H}_{\rm eff} &= -{4 G_F \over \sqrt{2}} V_{ts}^* V_{tb} \sum_i (C_i^\ell O_i^\ell +
                     C_i^{\prime \ell} O_i^{\prime \ell})+h.c.,
\end{align}
where $O_i^{(\prime)}$'s are dimension 5 and 6 $b \to s$ transition operators, for example,
\begin{align}
O_7 &= {e \over 16 \pi^2} m_b (\bar{s} \sigma^{\mu\nu} P_R b) F_{\mu\nu}, \quad
O'_7 = {e \over 16 \pi^2} m_b (\bar{s} \sigma^{\mu\nu} P_L b) F_{\mu\nu}, \nl
O_9^\ell &= {e^2 \over 16 \pi^2} (\bar{s} \gamma_\mu P_L b)(\bar{\ell} \gamma^\mu \ell), \quad
O^{'\ell}_9 = {e^2 \over 16 \pi^2} (\bar{s} \gamma_\mu P_R b)(\bar{\ell} \gamma^\mu \ell), \nonumber\\
O^\ell_{10} &= {e^2 \over 16 \pi^2} (\bar{s} \gamma_\mu P_L b)(\bar{\ell} \gamma^\mu \gamma_5 \ell), \quad
O^{'\ell}_{10} = {e^2 \over 16 \pi^2} (\bar{s} \gamma_\mu P_R b)(\bar{\ell} \gamma^\mu \gamma_5 \ell).
\end{align}
Writing $C_i^\ell = C_i^{\rm SM}+C_i^{\ell,\rm NP}$, the SM contribution at $m_b$ scale is $C_7^{\rm SM} \simeq -0.294$,
$C_9^{\rm SM}\simeq 4.20$, $C_{10}^{\rm SM}\simeq -4.01$.
The global fits to the experimental data show that the strongest pull is obtained in the scenario with NP in $C_9$ only~\cite{Altmannshofer:2017fio}.
The best fit value is $C_{9}^\mu =-1.21$ with pull 5.2$\sigma$.

There are already many works incorporating NP contribution to explain the violation of LFU in $b \to s \ell^+ \ell^-$ 
decays with tree-level $Z'$ contributions~\cite{Chang:2013hba, Crivellin:2015lwa,Allanach:2015gkd,
  Boucenna:2016wpr,Boucenna:2016qad,Kawamura:2017ecz,
  GarciaGarcia:2016nvr,Ko:2017quv, Ko:2017yrd, Ko:2017lzd, King:2017anf,
  DiChiara:2017cjq,Alonso:2017bff,Bonilla:2017lsq,Ellis:2017nrp,
Alonso:2017uky,Tang:2017gkz,Datta:2017ezo,Chiang:2017hlj,Choudhury:2017qyt},
 with leptoquarks~\cite{Bauer:2015knc,Das:2016vkr,Becirevic:2016yqi,Sahoo:2016pet,Hiller:2016kry,Bhattacharya:2016mcc,Becirevic:2017jtw,
Cai:2017wry,Das:2017kfo},
and with loop-processes~\cite{Belanger:2015nma,Gripaios:2015gra,Hu:2016gpe,DAmico:2017mtc,Kamenik:2017tnu,Poh:2017tfo}.

In this paper we propose a NP model with local $U(1)_{L_\mu-L_\tau}$ symmetry to solve the $b \to s \mu^+\mu^-$ anomaly.
This model naturally breaks LFU between $e$ and $\mu$ because the $\U1mt$ gauge boson couples only to $\mu (\tau)$ but
not to $e$. The model was originally proposed by He, Joshi, Lew, and Volkas~\cite{He:1990pn,He:1991qd}.
Many variants of $U(1)_{L_\mu-L_\tau}$ model have been studied ever since: the $Z'$ contribution to
the muon $(g-2)$ discrepancy~\cite{Baek:2001kca}, $U(1)_{L_\mu-L_\tau}$-charged dark matter (DM)~\cite{Baek:2008nz},
predictions on neutrino parameters~\cite{Baek:2015mna}, very light $Z'$ contribution to the annihilations of
DM~\cite{Baek:2015fea}.
Especially the $\U1mt$ model has also been extended to include to explain $b \to s \mu^+ \mu^-$
anomaly, but in different context from our
model~\cite{Altmannshofer:2014cfa,Crivellin:2015mga,Altmannshofer:2015mqa,Arnan:2016cpy,Altmannshofer:2016jzy,Chen:2017usq}. 
Our model has $U(1)_{L_\mu-L_\tau}$-charged colored scalars which we call {\it squark} coupling to $s,b$-quarks, while
the ref.~\cite{Crivellin:2015mga} introduces vector-like quarks. The former has one-loop contribution to $b \to s
\mu^+\mu^-$, whereas the latter has tree-level contribution.
Our model has also a natural dark matter candidate.  It corresponds to the scenario with NP in $C_9$ only
mentioned above, which is obtained from the $Z'$-penguin diagrams. There is no box-diagram contribution at one-loop
level. In our model $C_9$ includes contribution from $b \to s$ transition which comes from quark-squark-DM Yukawa
interaction as well as contribution from $U(1)_{L_\mu-L_\tau}$ gauge interactions. Although the $b \to s$ transition is strongly constrained
by other quark FCNC processes such as $b \to s \gamma$ and $B_s-\bar{B}_s$, we can evade them easily in our scenario.

The paper is organized as follows. In section~\ref{sec:model}, we introduce our model. The section~\ref{sec:NP} presents
the results for NP contributions to $b\to s \mu^+\mu^-$, $b \to s\gamma$, and $B_s -\ol{B}_s$ mixing and shows that we
can accommodate both $b \to s \mu^+\mu^-$ anomaly and the correct relic density of DM in our universe. In the
section~\ref{sec:DM} we discuss DM phenomenology. We conclude in section~\ref{sec:concl}.

%

\section{The model}
\label{sec:model}
We introduce a local $\U1mt$ symmetry in addition to the SM gauge group. The second (third) generation left-handed lepton
doublet and right-handed singlet, $\ell_L^{\mu}, \mu_R$,  ($\ell_L^{\tau}, \mu_R$), are charged under $\U1mt$ with
charge $1 (-1)$. It is a well-known fact that the $\U1mt$ is anomaly-free even without extending the SM particle
content. We also introduce new particles which are charged under $\U1mt$, a Dirac fermion $N$, a colored $SU(2)_L$-doublet scalar
$\wt{q} \equiv (\wt{u}, \wt{d})^T$, and a singlet-scalar $S$. Since the new fermion $N$ is a Dirac particle, the theory is
free from gauge anomaly.
Their charge assignments are shown in Table~\ref{tab:particles}.

\begin{table}[th]
\begin{center}
\begin{tabular}{|c|c|c|c|}\hline
&\multicolumn{1}{|c|}{New fermion} & \multicolumn{2}{c|}{New scalars} \\\hline
                 & $N$             & $\tilde{q}$       & $S$ \\ \hline
 $SU(3)_C$ &  {\bf 1}         &  {\bf 3}            &  {\bf 1}   \\ \hline
 $SU(2)_L$ &  {\bf 1}         &  {\bf 2}            &  {\bf 1}   \\ \hline
 $U(1)_Y$ &  $0$              &  ${1 \over 6}$            &  $0$   \\ \hline
 $\U1mt$ & $Q$ & $-Q$   & $2Q$     \\\hline
\end{tabular}
\caption{Charge assignments of $N, \wt{q}$ and $S$ under the SM gauge group and $\U1mt$. We take $Q \not =0, \pm 1$. }
\label{tab:particles}
\end{center}
\end{table}

The Lagrangian is written as
\begin{align}
{\cal L} &= {\cal L}_{\rm SM}  -V -{1 \over 4} Z'_{\mu\nu} Z^{'\mu\nu} + \ol{N} (i \gamma^\mu D_\mu -M_N) N + (D_\mu \wt{q}^\dagger) (D^\mu \wt{q})  -
           m_{\wt{q}}^2 \wt{q}^\dagger \wt{q}  \nl
           &+ (D_\mu S^\dagger)(D^\mu S)  - m_S^2 S^\dagger S - \sum_{i=s,b} (y_L^i \ol{q}_L^i \wt{q} N + h.c.) 
             -({f \over 2} \ol{N^c} N S^\dagger + h.c.),
\label{eq:Lag}
\end{align}
where $Z'_{\mu\nu}\equiv \partial_\mu Z'_\nu -\partial_\nu Z'_\mu$ is the field strength tensor for $\U1mt$ gauge boson
$Z'$, $D_\mu$ is the covariant derivative, and $i$ represents the generation index. The $N^c$ is the charge conjugate state of $N$. 
The scalar potential $V$ including the SM Higgs parts can be written as
\begin{align}
V &=  \lambda_H \left(H^\dagger H-\frac{v^2}{2} \right)^2 + \lambda_S \left(S^\dagger S-\frac{v_S^2}{2} \right)^2 + \lambda_{HS}
 \left(H^\dagger H -\frac{v^2}{2}\right)\left(S^\dagger S -\frac{v_S^2}{2}\right) \nonumber\\
&+\lambda_{H\wt{q}} \left(H^\dagger H -\frac{v^2}{2}\right) \wt{q}^\dagger \wt{q}  
+ \lambda'_{H\wt{q}} \left(H^\dagger \wt{q}\right) \left( \wt{q}^\dagger H \right)
+\lambda_{S\wt{q}} \left(S^\dagger S -\frac{v_S^2}{2}\right) \wt{q}^\dagger \wt{q}+ \lambda_{\wt{q}} \left(\wt{q}^\dagger \wt{q}\right)^2,
\label{eq:potential}
\end{align}
where $H$ is the SM Higgs doublet, $v = \sqrt{2} \langle H^0 \rangle$ and $v_S =\sqrt{2} \langle S \rangle$ are vacuum
expectation values (VEVs) of the scalar fields.
The field $\wt{q} (= (\wt{u}, \wt{d})^T)$ and $N$ mediate the $\U1mt$ interaction to quark sector. Although the generation, $i=d$, is allowed by the gauge
symmetry in general, we neglect the interaction with the first generation in this paper, because it is irrelevant in our discussion of $b \to s
\mu\mu$ transition. We assume that the down-type quarks in (\ref{eq:Lag}) are already in the mass eigenstates and that the flavor mixing due to 
Cabibbo-Kobayashi-Maskawa (CKM) mixing appears
only in the up-quark sector, {\it i.e.} $d_L=d'_L, u_L =V_{\rm CKM}^\dagger u'_L$ with $d'_L,u'_L$ being the mass
eigenstates. 
There is mass splitting between $\wt{u}$ and $\wt{d}$ due to $\lambda'_{H\wt{q}}$ term:
\begin{align}
m^2_{\wt{u}} &= m^2_{\wt{q}}, \nl
m^2_{\wt{d}} &= m^2_{\wt{q}} + {1 \over 2} \lambda'_{H\wt{q}} v^2.
\end{align}

The flavored-DM scenarios where DMs have interactions with quarks or leptons in a form in (\ref{eq:Lag}) 
have been also studied in~\cite{Baek:2015fma,Baek:2016kud,Baek:2016lnv,Kawamura:2017ecz}.
Since $N$ and $N^c$ are mixed after $S$ gets VEV, they are not mass eigenstates. The mass matrix of $N$ and $N^c$ is written as
\begin{align}
\left(
\begin{array}{cc}
M_N & \frac{f v_S}{\sqrt{2}} \\
\frac{f v_S}{\sqrt{2}} & M_N
\end{array}
\right).
\end{align}
The mass eigenstates are  mixture of $N$ and $N^c$,
\begin{align}
N_- &= {1 \over \sqrt{2}} (N- N^c), \nl
N_+ &= {1 \over \sqrt{2}} (N+ N^c),
\end{align}
whose masses are $m_\mp =M_N \mp f v_S/\sqrt{2}$.  The $N_\mp$ are two Majorana particles.
The $N_-$ state has Majorana phase $\pi$ so that $N_-^c = - N_-$, but $N_+^c =N_+$.
In the unitary gauge, $S$ can be decomposed as $S =1/\sqrt{2}(v_S + \phi_S)$. The real scalar $\phi_S$ can also mix with
the SM Higgs boson $h$ via $\lambda_{HS}$ term in (\ref{eq:potential}). The mixing angle is denoted as $\alpha_H$. 
The size of the mixing is constrained by the LHC
experiments. The details can be found in~\cite{Baek:2011aa, Baek:2012se}. In this work, the mixing between $\phi_S$ and
$h$ does not affect $b \to s \mu^+ \mu^-$, but it affects the dark matter phenomenology. The direct detection 
experiments of dark matter strongly constrains this {\it Higgs portal interaction}. We take 
$\alpha_H \le 0.1$ in our numerical analysis to evade the constraints from LHC experiments. We will see that it is
further constrained by the DM direct detection experiments.

After $\U1mt$ is broken by the VEV of $S$, $v_S$, there is still remnant $Z_2$ symmetry due to the last terms in
(\ref{eq:Lag}).
This discrete symmetry stabilizes the lightest neutral $Z_2$ odd particle, which we assume is $N_-$.
And it becomes a good dark matter candidate. As mentioned above the DM interacts with the SM sector mediated either by
the Higgs portal or $Z'$.  The DM pair annihilates through $N_- N_- \to Z' Z'$.  The coannihilation
process $N_- N_+ \to Z' \to \mu^+ \mu^- (\tau^+ \tau^-)$ is possible in case the mass difference $\Delta m\equiv m_+ -m_- =\sqrt{2}
f v_S$ is small. There is also annihilation into the SM particles through the $\phi_S$ and the SM Higgs mixing (Higgs
portal).  Notice that we take the $\U1mt$ charge of $N$, $Q\not= 0, \pm 1$, so that
the Yukawa interactions $\ol{\ell_L^\mu} \wt{H} N$ or $\ol{\ell_L^\tau} \wt{H} N$ are {\it not} allowed. Otherwise $N$ can
mix with the active neutrinos, $\nu_\mu$ or $\nu_\tau$, and $N$ cannot be the electroweak scale WIMP DM candidate.

After $S$ gets VEV, the $\U1mt$ gauge boson gets mass, $m_{Z'} = 2 g_X |Q| v_S$, where $g_X$ is the $\U1mt$ gauge coupling
constant. 
The squarks $\wt{u}$ and $\wt{d}$ have masses $m^2_{\wt{u}}=m_{\wt{q}}^2$ and $m^2_{\wt{d}}=m_{\wt{q}}^2+\lambda'_{Hq}
v^2/2$, respectively. Since large mass splitting leads large contribution to $\rho$-parameter~\cite{Patrignani:2016xqp}
and the mass splitting does not affect the analysis, we set $\lambda'_{Hq}=0$ for simplicity.

\section{NP contribution to $b \to s$ transitions in our model}
\label{sec:NP}

\begin{figure}[tbp]
\begin{center}
\includegraphics[width=.85\textwidth]{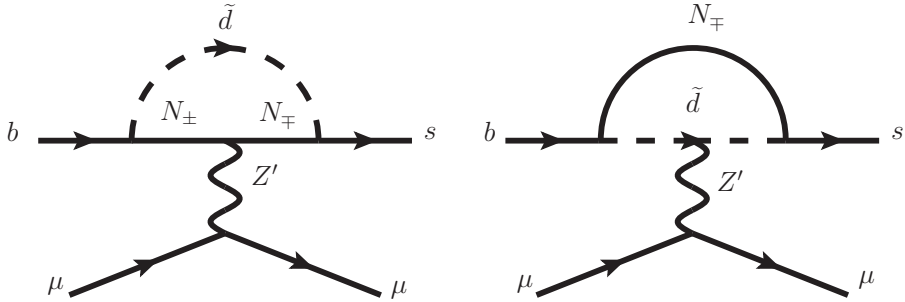}
\end{center}
\caption{ Penguin diagrams generating $b \to s \mu^+ \mu^-$ in our model.}
\label{fig:O9}
\end{figure}

\begin{figure}[tbp]
\begin{center}
\includegraphics[width=.45\textwidth]{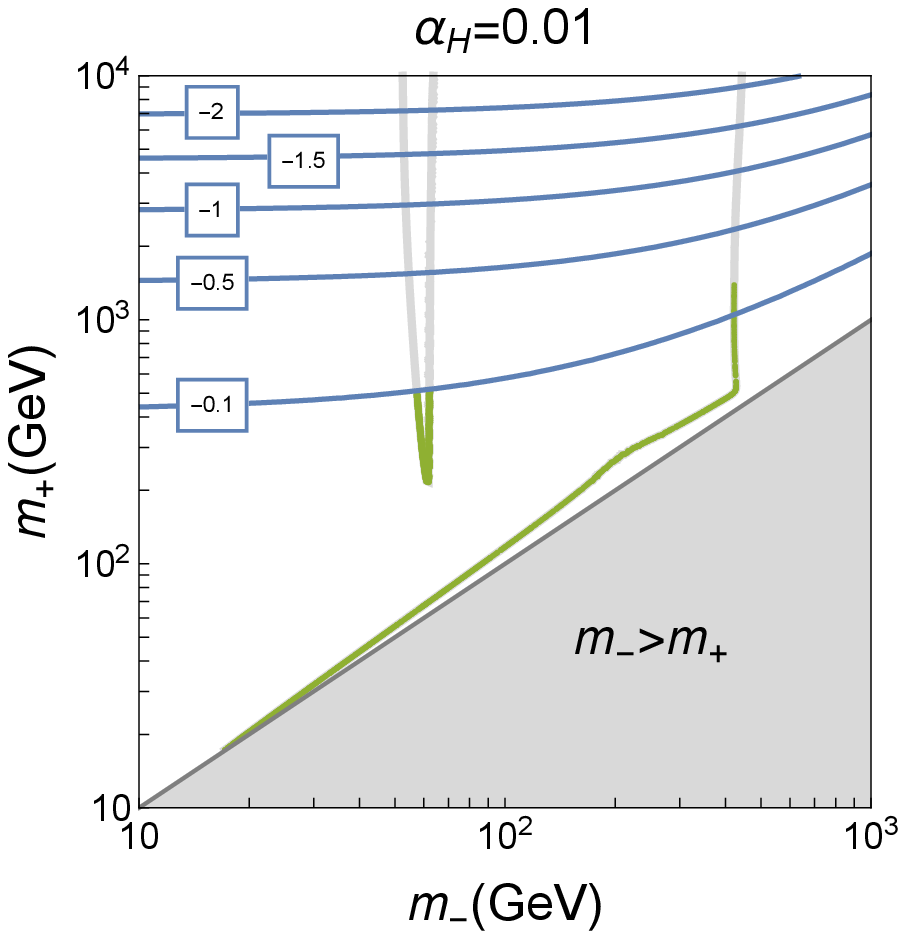}
\includegraphics[width=.45\textwidth]{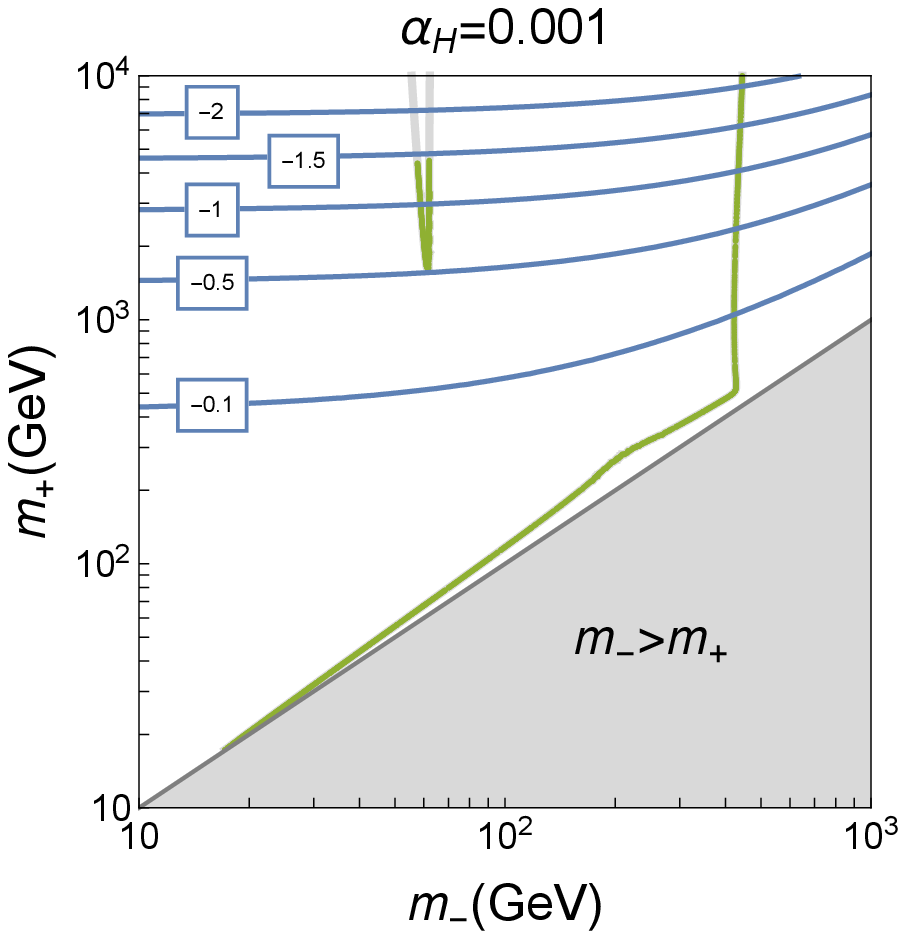}
\end{center}
\caption{Contour plot for $C_9^{\mu,{\rm NP}}$ at the electroweak scale (blue lines) and $\Omega h^2 =0.12$ (green and grey lines)
  in the  plane $(m_-,m_+)$.   We fixed $Q=3/2$, $\alpha_X=0.05$, $y_L^s y_L^b=0.1$,
$m_{Z'}=500$ GeV, $m_{\wt{d}}=3$ TeV, $m_{H_1}=125$ GeV, and $m_{H_2}=1$ TeV. We also set $\alpha_H$=0.01 (left panel)
and $\alpha_H$=0.001 (right panel). The grey parts of the lines are excluded by the DM direct detection experiments.}
\label{fig:cont_C9}
\end{figure}

In our model the $b \to s \mu^+ \mu^-$ transition operator $O_9^\mu$ is generated by $Z'$-exchanging penguin diagrams as
shown in Fig.~\ref{fig:O9}. From the diagrams we can see that only $C_9^{\mu(\tau),{\rm NP}}$ is generated. There is no
box diagram contribution at one-loop level. The result is
\begin{align}
C_9^{\mu,{\rm NP}} &= -Q \frac{\sqrt{2}}{4 G_F m_{Z'}^2} \frac{\alpha_X}{\alpha_{\rm em}} \frac{y_L^s y_L^b }{V_{ts}^*
                     V_{tb}} {\cal V}_{sb}(x_-,x_+),
\end{align}
where $\alpha_X = g_X^2/(4 \pi)$, $x_\mp = m^2_\mp/m^2_{\wt{d}}$, and ${\cal V}_{tb}(x_-,x_+)$ is the 
effective $b-s-Z'$ vertex at zero momentum transfer. The
expression for ${\cal V}_{sb}$ is given in the appendix~\ref{app:loop}. In the limit $m_- = m_+$ we obtain
${\cal  V}_{sb}(x_-,x_-)=0$. This can be understood from the fact we took zero momentum limit to get ${\cal V}_{sb}$. In the limit 
$m_-=m_+$, the two Majorana fermions $N_\mp$ return to the original Dirac fermion $N$. Note $\Delta m= \sqrt{2} f v_S=0$
corresponds to $v_S=0$. And the $\U1mt$ gauge symmetry is
restored, and ${\cal V}_{sb}(x_-,x_-)=0$ is simply a consequence of $\U1mt$ gauge invariance. 
Therefore, to get a sizable $C_9^{\mu,{\rm NP}}$ we need large mass splitting $\Delta m$, which implies the DM
coannihilation processes for the DM relic cannot be a dominant component  in our scenario.
Fig.~\ref{fig:cont_C9} shows a contour plot for $C_9^{\mu,{\rm NP}}$ (blue lines) in the plane $(m_-,m_+)$. For the
plot we fixed $Q=3/2$, $\alpha_X=0.05$, $y_L^s y_L^b=0.1$,  $m_{Z'}=500$ GeV, $m_{H_1}=125$ GeV, , $m_{H_2}=1$ TeV, and
$m_{\wt{d}}=3$ TeV. The left (right) panel corresponds to $\alpha_H=0.01 (0.001)$. This choice of $\alpha_X$ and
$m_{Z'}$ can be target of LHC searches for $Z'$~\cite{Altmannshofer:2016jzy}.
In the same plot we also show $\Omega h^2 =0.12$ lines (green and gray lines),
which could explain the DM relic in the universe. The green (gray) parts are (not) allowed by the DM direct search
experiments. We used the upper limit from LUX for the plot~\cite{Akerib:2016vxi}.
The DM phenomenology will be discussed in more detail in Section~\ref{sec:DM}. 
We can see that $C_9^{\mu,{\rm NP}} \sim -1$ can be obtained with sizable mass splittings in the case
$\alpha_H=0.001$, which can explain the $b \to \mu^+\mu^-$ anomaly. Smaller $\alpha_H$ is preferred because
it can help evade the direct search bound whereas the scalar mixing does not affect $C_9^{\mu,{\rm NP}}$. 
We also note that the value of $C_9^{\mu,{\rm NP}}$ does not depend on the
absolute value of $m_{\wt{d}}$ or $m_\mp$ but only on their ratios. We take $m_{\wt{d}}=3$ TeV as a reference value.


\begin{figure}[tbp]
\begin{center}
\includegraphics[width=.5\textwidth]{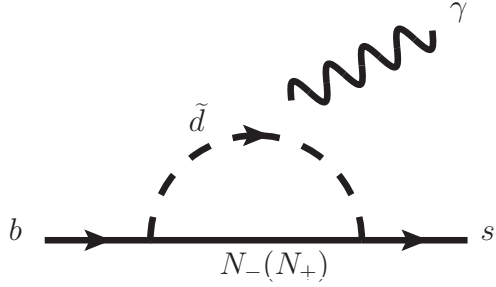}
\end{center}
\caption{Feynman diagrams for $b \to s \gamma$ in our model. The photon can be attached to electrically charged particles, $\wt{d}, b$ or $s$. }
\label{fig:bsr}
\end{figure}

The couplings $y_L^s$ and $y_L^b$ also generate other quark FCNC processes such as $b \to s \gamma$ and $B_s-\ol{B_s}$
mixing. The experimental measurement of the inclusive branching fraction of radiative $B$-decay, $\ol{B} \to X_s \gamma$,
is~\cite{Amhis:2016xyh} 
\begin{align}
{\cal B}\left[\ol{B} \to X_s \gamma, \left(E_\gamma > {1 \over 20} m_b\right) \right]^{\rm exp} &= (3.32 \pm 0.15)
                                                                                                  \times 10^{-4}, 
\end{align}
which can be compared with theoretical prediction in the SM prediction~\cite{Misiak:2015xwa}
\begin{align}
{\cal B}\left[\ol{B} \to X_s \gamma, \left(E_\gamma > 1.6 \,{\rm GeV}\right) \right]^{\rm SM} &= 
(3.36 \pm 0.23)  \times 10^{-4} .
 \end{align}
The NP contribution to $C_7^\gamma$ at the electroweak scale whose diagram is shown in Fig.~\ref{fig:bsr} is obtained to be
\begin{align}
C_7^{\gamma, {\rm NP}}&= \frac{\sqrt{2} e_d}{16 G_F m^2_{\wt{d}}} \frac{y_L^s y_L^b}{V_{ts}^* V_{tb}} [J_1(x_-)+J_1(x_+)],
\end{align}
where $e_d=-1/3$ is the electric charge of $\wt{d}$, and the loop function $J_1(x)$ is given in the Appendix~\ref{app:loop}.
The corresponding SM value is $C_7^{\gamma,{\rm SM}}(\mu_b) \simeq -0.294$~\cite{Buchalla:1995vs}.

\begin{figure}[tbp]
\begin{center}
\includegraphics[width=.85\textwidth]{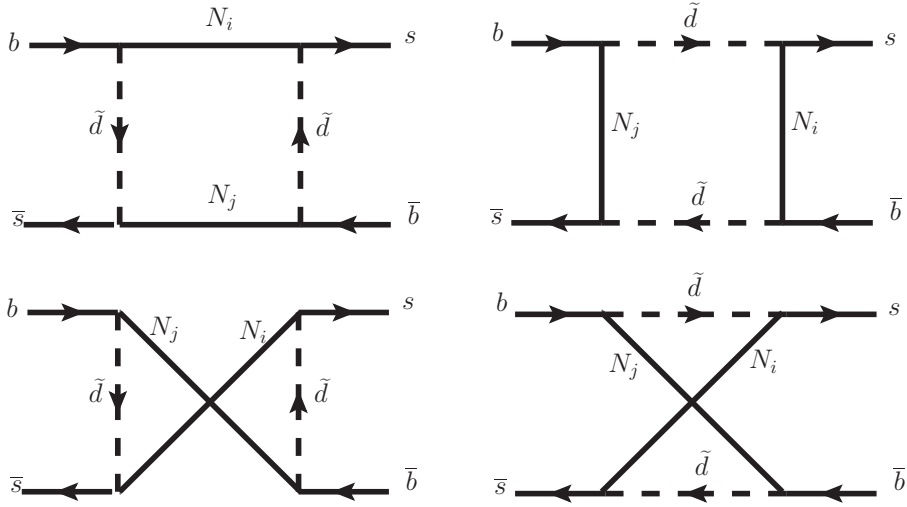}
\end{center}
\caption{Box diagrams which contribute to $B_s-\ol{B}_s$ mixing, where $i,j=\pm$.}
\label{fig:BsBs}
\end{figure}

The NP contribution to $B_s-\ol{B}_s$ mixing occurs via the box diagrams shown in Fig.~\ref{fig:BsBs}. The mass difference in the
$B_s-\ol{B}_s$ system has been measured by CDF and LHCb and the average value is
\begin{align}
\Delta m_s &=17.757 \pm 0.020 (\rm stat) \pm 0.007 (\rm syst) \; {\rm ps}^{-1},
\end{align}
which is in good agreement with the SM predictions~\cite{Olive:2016xmw} with predictions $17.5 \pm 1.1 \; {\rm ps}^{-1}$~\cite{Bona:2006ah} or 
$16.73^{+0.82}_{-0.57} \; {\rm ps}^{-1}$~\cite{Charles:2015gya}. More recently the ref.~\cite{Jubb:2016mvq} reports larger central value for
the SM prediction but with larger errors, $\Delta m_s^{\rm SM} = 18.6^{+2.4}_{-2.3}$. 
The effective Hamiltonian in the SM has $(V-A) \times
(V-A)$ structure since the $W$-boson couples only to left-handed quarks,
\begin{align}
{\cal H}_{\rm eff}^{\Delta B=2} &= \frac{G_F m_W^2}{4 \pi^2} (V_{ts}^* V_{tb})^2 S_0(x_t) 
           \ol{s} \gamma_\mu P_L b \ol{s}  \gamma^\mu P_L b 
 \equiv C_1^{\rm SM}(\mu_W)    \ol{s} \gamma_\mu P_L b \ol{s}  \gamma^\mu P_L b ,
\label{eq:BsBsSM}
\end{align}
where $x_t = m_t^2/m_W^2$ and the loop function $S_0(x_t)$ can be found, {\it e.g.}, in~\cite{Buchalla:1995vs}.
The NP contribution to the $B_s-\ol{B}_s$ mixing whose Feynman diagrams are shown in Fig.~\ref{fig:BsBs}  can also be written
in the form
\begin{align}
{\cal H}_{\rm eff}^{\Delta B=2, {\rm NP}} &= C_1^{\rm NP} \ol{s} \gamma_\mu P_L b \ol{s} \gamma^\mu P_L b,
\end{align}
where at the electroweak scale
\begin{align}
C_1^{\rm NP}(\mu_W) &=\frac{(y_L^s y_L^b)^2}{128 \pi^2 m^2_{\wt{d}}} \Big[ 2 k(x_-,x_-,1) +4 k(x_-,x_+,1) +
                   2   k(x_+,x_+,1)  \nl
 &+ x_- j(x_-,x_-,1) + 2 \sqrt{x_- x_+} j(x_-,x_+,1) +x_+ j(x_+,x_-,1)
                       \Big],
\end{align}
where the loop functions $j$ and $k$ are listed in the Appendix~\ref{app:loop}.
Since new particles in our model couple only to the left-handed quarks as shown in (\ref{eq:Lag}), the NP operator has
the same Lorentz structure with the SM operator in (\ref{eq:BsBsSM}).

By comparing the experimental results with the SM predictions, we 
can see that both $b \to s \gamma$ and $B_s-\ol{B}_s$ mixing still allows $\sim 10$ \% contribution from NP.
To see the impact of these FCNC constraints on our model, we show contour plots for 
$C_7^{\gamma, {\rm NP}}/C_7^{\gamma,    {\rm SM}}$ for $b \to s \gamma$
and $C_1^{\rm NP}/C_1^{\rm SM}$ for $B_s-\ol{B}_s$ mixing at the electroweak scale. We take the same parameters
with Fig.~\ref{fig:cont_C9}. We see the NP contribution to the $b \to s \gamma$ is typically a few$\times O(10^{-4})$ of
the SM contribution and its contribution to the $B_s-\ol{B}_s$ mixing is a few \% near the region where $C_9^{\mu,{\rm
    NP}} \sim -1$. Consequently we can
safely evade the constraints for the parameters chosen in Fig.~\ref{fig:cont_C9} while getting sizable 
$C_9^{\mu,{\rm   NP}} \sim -1$ to resolving the $b \to s \mu\mu$ anomaly. We also checked even for the relatively
large $\Delta m=\sqrt{2} f v_S$, the dark Yukawa coupling $f$ can be still in the perturbative regime, {\it i.e.} $f \lesssim 4 \pi$.   
The key observation is that the large
contribution to $C_9^{\mu,{\rm NP}}$ comes from the
relatively light $Z'$ ($m_{Z'} \sim O(100)$ GeV) and sizable $\U1mt$ coupling, $\alpha_X \sim 0.1$, while $Z'$ gauge
boson is not involved in $b\to s\gamma$ or $B_s-\ol{B}_s$ mixing at the one-loop level.

Our model can contribute also possibly to $B \to K^{(*)} \nu \bar{\nu}$ processes. The leading diagrams are obtained by
replacing $\mu$ with $\nu_\mu$ and $\nu_\tau$ in Fig.~\ref{fig:O9}. However, since their $\U1mt$ charges, $Q_\mu =+1$ 
and $Q_\tau=-1$, add to zero, the two contributions cancel with each other. We predict there is no deviation from the SM
predictions in $B \to K^{(*)} \nu \bar{\nu}$ decays.

\begin{figure}[tbp]
\begin{center}
\includegraphics[width=.5\textwidth]{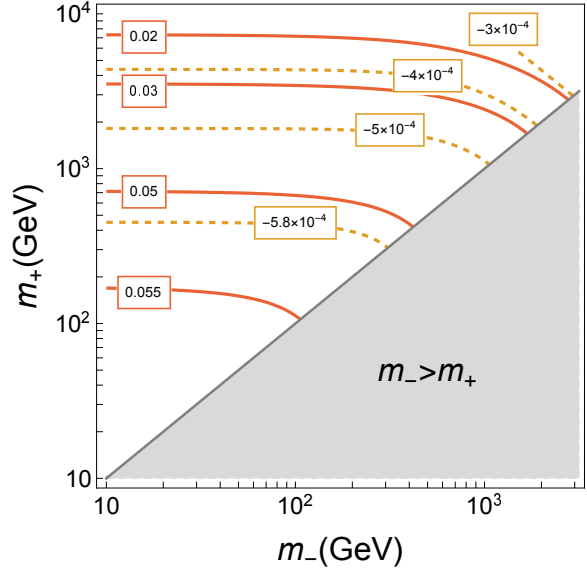}
\end{center}
\caption{Contour plots for $C_7^{\gamma,{\rm NP}}/C_7^{\gamma,{\rm SM}}$ (dashed lines) and 
$C_1^{\rm NP}/C_1^{\rm SM}$ (solid lines) at the electroweak scale. The fixed parameters are the same with Fig.~\ref{fig:cont_C9}.}
\label{fig:constr}
\end{figure}

\section{Dark matter relic density and direct detection}
\label{sec:DM}

\begin{figure}[tbp]
\begin{center}
\includegraphics[width=.5\textwidth]{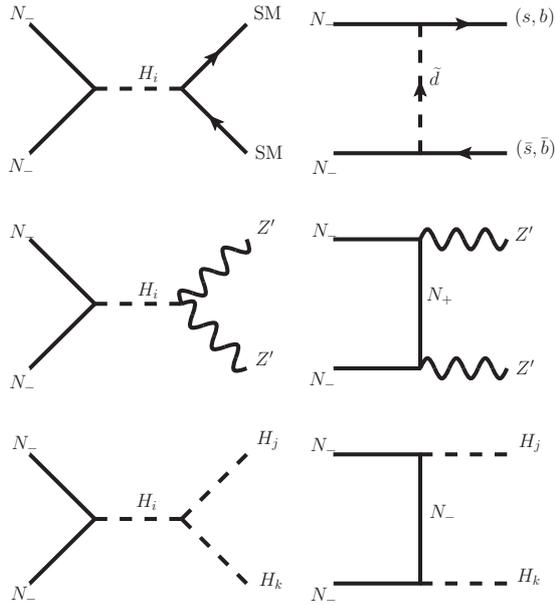}
\end{center}
\caption{DM annihilation diagrams for relic density.}
\label{fig:ann}
\end{figure}

There are many DM annihilation diagrams for the relic density in our model: $H_1 (H_2)$-mediated $s-$channel diagrams
for $N_- N_- \to {\rm SM\; SM}$ (Higgs portal contributions), $\wt{d}-$mediated $t-$channel diagrams for 
$N_- N_- \to s s, s b, b b$, and $N_- N_- \to Z' Z', H_j H_k (j,k=1,2)$  which have both
Higgs-mediated $s-$channel and $N_+ (N_-)-$mediated $t-$channel diagrams. They are shown in Fig.~\ref{fig:ann}.

Also coannihilation diagrams can make contributions when 
$\Delta m (=m_+-m_-) \approx m_-/20$. 
To calculate the relic density and DM nucleon scattering cross section we implemented our model to the numerical
package micrOMEGAs~\cite{Belanger:2014vza}.

Fig.~\ref{fig:cont_C9} shows contour lines for the total DM relic density $\Omega_{\rm DM} h^2 \approx 0.12$ (green and
gray lines). We can clearly see three main contributions: i) we can see Higgs resonance contribution dominating near 
$m_- \approx m_{H_1}/2 \approx 65$ GeV, ii) the coannihilation contributions are important along the line $m_+ \approx m_-$, iii) $N_- N_- \to Z' Z'$
process takes over when it is kinematically open near $m_- \gtrsim m_{Z'}$.
The green (gray) parts satisfy (do not satisfy) the constraints from the DM and nucleon scattering
experiments.
When the coannihilation dominates, it is difficult to get sizable
$C_9^{\mu,{\rm NP}}$, which can be seen also in Fig.~\ref{fig:cont_C9}.
In the right panel the green lines with $C_9^{\mu,{\rm NP}} \sim -1$ intersect with the lines with $\Omega h^2 =0.12$, while
satisfying the direct detection constraints. And we can accommodate both the $R_{K^{(*)}}$
anomaly and the correct DM relic density of the universe in our model.

The main contribution to the direct detection experiments comes from the $H_i (i=1,2)$-exchanging $t$-channel diagrams
(Higgs portal). Since the Higgs portal contribution favors large $\alpha_H$, it is clear to see that the Higgs-resonance region
is more strongly constrained. 
By the same reason the constraint is more stringent in the left panel than the right.
We note that to explain both the $B$-anomaly and the null search of DM with nucleon scattering the bound on $\alpha_H$
becomes much more stronger $\alpha_H \lesssim 0.01$ than the collider bound $\alpha_H \lesssim 0.1$.

\section{Conclusions}
\label{sec:concl}

The LHCb and Belle experiments have observed tantalizing anomalies in $b \to s \mu^+ \mu^-$ processes
in the observables of $P_5', R_{K^{(*)}}$ and some branching fractions. The global fits show $\sim 5
\sigma$ deviation from the standard model predictions with the best fit value $C_9^{\mu,{\rm NP}} \sim -1$ for
$C_9^{\mu,\rm NP}$ only scenario.
We propose a local $\U1mt$ model which correspond to this scenario.
The model also contains a natural dark matter candidate. The new physics contribution to $b \to s$
transition occurs via the exchange of colored $SU(2)_L$-doublet scalar $\wt{q}$ which is also charged under $\U1mt$. To
conserve $\U1mt$ charge in the Yukawa interaction of $\wt{q}$ with the SM quarks $b,s$, we need a 
Dirac fermion $N$ which is electrically neutral but charged under $\U1mt$
so that the Yukawa interaction takes the form  $y_L^i \wt{q}^i_L \wt{q} N + h.c.$. The neutral fermion can be a dark
matter candidate.
The stability of dark matter candidate $N$ is
achieved by the interaction $f \ol{N^c} N S^\dagger + h.c.$ where $S$ is $\U1mt$-breaking scalar, which leaves exact
$Z_2$ symmetry after $S$ obtains vacuum expectation value. The lightest $Z_2$-odd neutral Majorana fermion mass eigenstate
$N_-$ becomes a stable dark matter.

The model contributes to the $C_9^\mu$ for $b \to s \mu^+ \mu^-$ through the $Z'$-exchanging penguin
diagrams. When $\alpha_X \sim 0.05$, $m_{Z'} \sim 500$ GeV, we can explain the $R_{K^{(*)}}$ anomaly since our
$Z'$ does not couple to electrons but to muons,  resulting in the violation of lepton flavor universality. Large mass splitting between $N_\pm$ states are
favored for large $C_9^{\mu,{\rm NP}}$.
The constraints on $b \to s$ transition from $b
\to s \gamma$ and $B_s-\ol{B}_s$ mixing can be evaded by taking relatively heavy ($\sim 3$ TeV) $\wt{q}$ and sizable
product of Yukawa couplings, $y_L^s y_L^b \sim 0.1$.
We predict that there is no deviation from the SM predictions in $B \to K^* \nu \ol{\nu}$ process, which can be tested at Belle-II. 

Our dark matter can provide the correct relic density via the annihilations of the Higgs resonance channel and $N_- N_-
\to Z' Z'$ channel. The latter channel is naturally realized in our model because $Z'$ is relatively light and has
sizable gauge coupling with the dark fermions $N_\pm$.
Our model can be tested by searching for $Z'$ and new colored scalar at the LHC.

\appendix
\section{Loop functions}
\label{app:loop}
The effective $b-s-Z'$ vertex at zero momentum transfer is given by the loop function at zero momentum transfer,
\begin{align}
{\cal V}_{sb}(x_-,x_+) &=\frac{1}{4}+\sqrt{x_- x_+} j(x_-,x_+)-\frac{1}{2} k(x_-,x_+)+I(x_-)+I(x_+),
\end{align}
where $x_\mp = m^2_\mp/m^2_{\wt{d}}$ and
\begin{align}
j(x) &= \frac{ x \log x}{x-1}, \nl
k(x) &= \frac{ x^2 \log x}{x-1}, \nl
I(x) &= \frac{-3x^2+4x-1+2x^2 \log x}{8(x-1)^2}.
\label{eq:lf_c9}
\end{align}
The loop functions of $j$ and $k$ with more than one argument are defined recursively as 
\begin{align}
f(x_1,x_2, x_3, \cdots) \equiv \frac{f(x_1,x_3, \cdots) - f(x_2, x_3, \cdots)}{x_1 - x_2}, 
\end{align}
where $f=j,k$.
These multi-argument functions of $j$ and $k$ also appear as loop functions for the $\Delta B=2$ box diagrams.

The loop function $J_1(x)$ for $b \to s \gamma$ is obtained to be
\begin{align}
J_1(x) &=\frac{1-6x+3x^2+2x^3-6x^2 \log x}{12(1-x)^4}.
\end{align}

\acknowledgments
The author is grateful to Yuji Omura, Jusak Tandean and Chaehyun Yu for useful discussions.
This work is supported in part by National Research Foundation of Korea (NRF) Research Grant NRF-2015R1A2A1A05001869.

\bibliographystyle{JHEP}
\bibliography{RK}

\end{document}